\documentclass[twocolumn]{aastex631}

\usepackage{graphicx}
\usepackage{amsmath}
\usepackage{amssymb}
\usepackage{subfigure}
\usepackage{cases}
\usepackage{mathrsfs}
\usepackage[flushleft]{threeparttable}
\usepackage{amssymb}
\usepackage{pifont}
\usepackage{epstopdf}
\usepackage{xcolor}
\usepackage[all]{hypcap}
\usepackage{mwe,tikz}
\usepackage{bm}
\usepackage{tabularx}
\usepackage{multirow}
\usepackage{longtable}
\usepackage{xspace}
\usepackage{rotating}
\usepackage{tikz-imagelabels}

\newcommand{\e}{\text{e}}

\newcommand{\p}{\partial}

\imagelabelset{
coarse grid color = red,
fine grid color = gray,
image label font = \sffamily\bfseries\small,
image label distance = 2mm,
image label back = white,
image label text = black,
coordinate label font = \sffamily\bfseries\scriptsize,
coordinate label distance = 2mm,
coordinate label back = white,
coordinate label text = black,
annotation font = \normalfont\small,
arrow distance = 1.5mm,
border thickness = 0.6pt,
arrow thickness = 0.4pt,
tip size = 1.2mm,
outer dist = 0.5cm,
}
\usepackage{CJK}
\usepackage{upgreek}
\usepackage{booktabs}

\begin{document}

\title{GRB 220408B: A Three-Episode Burst from a Precessing Jet}

\correspondingauthor{Bin-Bin Zhang, Ming Zeng, Hua Feng}
\email{bbzhang@nju.edu.cn,zengming@tsinghua.edu.cn,hfeng@tsinghua.edu.cn}

\author[0000-0002-2420-5022]{Zijian Zhang}
\affiliation{School of Astronomy and Space Science, Nanjing University, Nanjing 210093, China}

\author[0000-0002-5596-5059]{Yihan Yin}
\affiliation{School of Physics, Nanjing University, Nanjing 210093, China}

\author{Chenyu Wang}
\affiliation{Department of Astronomy, Tsinghua University, Beijing 100084, China}

\author[0000-0002-9738-1238]{Xiangyu Ivy Wang}
\affiliation{School of Astronomy and Space Science, Nanjing University, Nanjing 210093, China}
\affiliation{Key Laboratory of Modern Astronomy and Astrophysics (Nanjing University), Ministry of Education, China}

\author[0000-0002-5485-5042]{Jun Yang}
\affiliation{School of Astronomy and Space Science, Nanjing University, Nanjing 210093, China}
\affiliation{Key Laboratory of Modern Astronomy and Astrophysics (Nanjing University), Ministry of Education, China}

\author{Yan-Zhi Meng}
\affiliation{School of Astronomy and Space Science, Nanjing University, Nanjing 210093, China}
\affiliation{Key Laboratory of Modern Astronomy and Astrophysics (Nanjing University), Ministry of Education, China}

\author{Zi-Ke Liu}
\affiliation{School of Astronomy and Space Science, Nanjing University, Nanjing 210093, China}
\affiliation{Key Laboratory of Modern Astronomy and Astrophysics (Nanjing University), Ministry of Education, China}

\author{Guo-Yin Chen}
\affiliation{School of Astronomy and Space Science, Nanjing
University, Nanjing 210093, China}
\affiliation{Key Laboratory of Modern Astronomy and Astrophysics (Nanjing University), Ministry of Education, China}

\author{Xiaoping Fu}
\affiliation{School of Information Engineering, Nanchang University, Nanchang 330031, China}

\author[0000-0001-8082-893X]{Huaizhong Gao}
\affiliation{Department of Engineering Physics, Tsinghua University, Beijing 100084, China}
\affiliation{Department of Health Physics, China Institute for Radiation Protection, Taiyuan 300000, China}

\author[0000-0002-7796-5578]{Sihao Li}
\affiliation{Key Laboratory of Radiation Physics and Technology of Ministry of Education, College of Physics of Sichuan University, Chengdu, China}

\author[0000-0002-7619-6017]{Yihui Liu}
\affiliation{Department of Engineering Physics, Tsinghua University, Beijing 100084, China}

\author[0000-0002-1590-9327]{Xiangyun Long}
\affiliation{Department of Engineering Physics, Tsinghua University, Beijing 100084, China}

\author{Yong-Chang Ma}
\affiliation{School of Artificial Intelligence, Nanjing University, Nanjing 210093, China}

\author[0000-0002-7439-6621]{Xiaofan Pan}
\affiliation{Key Laboratory of Particle and Radiation Imaging (Tsinghua University), Ministry of Education, Beijing 100084, China}
\affiliation{Department of Engineering Physics, Tsinghua University, Beijing 100084, China}

\author[0000-0003-0080-1767]{Yuanze Sun}
\affiliation{Department of Computer Science and Technology, Nanjing University, Nanjing 210093, China}

\author{Wei Wu}
\affiliation{School of Information Engineering, Nanchang University, Nanchang 330031, China}

\author[0000-0003-0877-4345]{Zirui Yang}
\affiliation{Key Laboratory of Particle and Radiation Imaging (Tsinghua University), Ministry of Education, Beijing 100084, China}
\affiliation{Department of Engineering Physics, Tsinghua University, Beijing 100084, China}

\author[0000-0001-6438-7986]{Zhizhen Ye}
\affiliation{Key Laboratory of Radiation Physics and Technology of Ministry of Education, College of Physics of Sichuan University, Chengdu, China}

\author{Xiaoyu Yu}
\affiliation{School of Chemistry and Chemical Engineering , Nanjing
University, Nanjing 210093, China}

\author{Shuheng Zhao}
\affiliation{College of Engineering and Applied Sciences , Nanjing University, Nanjing 210093, China}

\author[0000-0001-7647-7110]{Xutao Zheng}
\affiliation{Key Laboratory of Particle and Radiation Imaging (Tsinghua University), Ministry of Education, Beijing 100084, China}
\affiliation{Department of Engineering Physics, Tsinghua University, Beijing 100084, China}

\author[0000-0001-6792-5145]{Tao Zhou}
\affiliation{Key Laboratory of Radiation Physics and Technology of Ministry of Education, College of Physics of Sichuan University, Chengdu, China}

\author[0000-0001-7471-8451]{Qing-Wen Tang}
\affiliation{Department of Physics, School of Physics and Materials Science, Nanchang University, Nanchang 330031, China}

\author[0000-0003-4736-7435]{Qiurong Yan}
\affiliation{School of Information Engineering, Nanchang University, Nanchang 330031, China}

\author[0000-0001-6399-8566]{Rong Zhou}
\affiliation{Key Laboratory of Radiation Physics and Technology of Ministry of Education, College of Physics of Sichuan University, Chengdu, China}

\author[0000-0002-6351-4696]{Zhonghai Wang}
\affiliation{Key Laboratory of Radiation Physics and Technology of Ministry of Education, College of Physics of Sichuan University, Chengdu, China}

\author{Hua Feng}
\affiliation{Department of Astronomy, Tsinghua University, Beijing 100084, China}

\author{Ming Zeng}
\affiliation{Key Laboratory of Particle and Radiation Imaging (Tsinghua University), Ministry of Education, Beijing 100084, China}
\affiliation{Department of Engineering Physics, Tsinghua University, Beijing 100084, China}

\author[0000-0003-4111-5958]{Bin-Bin Zhang}
\affiliation{School of Astronomy and Space Science, Nanjing
University, Nanjing 210093, China}
\affiliation{Purple Mountain Observatory, Chinese Academy of Sciences, Nanjing 210023, China}
\affiliation{Key Laboratory of Modern Astronomy and Astrophysics (Nanjing University), Ministry of Education, China}

% \author[0000-0003-4111-5958]{Someone Else}
% \affiliation{School of Astronomy and Space Science, Nanjing
% University, Nanjing 210093, China}
% \affiliation{Key Laboratory of Modern Astronomy and Astrophysics (Nanjing University), Ministry of Education, China}

\begin{abstract} 
 
Jet precession has previously been proposed to explain the apparently repeating features in the light curves of a few gamma-ray bursts (GRBs). In this {\it Letter}, we further apply the precession model to a bright GRB 220408B by examining both its temporal and spectral consistency with the predictions of the model. As one of the recently confirmed GRBs observed by our GRID CubeSat mission, GRB 220408B is noteworthy as it exhibits three apparently similar emission episodes. Furthermore, the similarities are reinforced by their strong temporal correlations and similar features in terms of spectral evolution and spectral lags. Our analysis demonstrates that these features can be well explained by the modulated emission of a Fast-Rise-Exponential-Decay (FRED) shape light curve intrinsically produced by a precessing jet with a precession period of $18.4 \pm 0.2$ seconds, a nutation period of $11.1 \pm 0.2$ seconds and viewed off-axis. This study provides a straightforward explanation for the complex yet similar multi-episode GRB light curves.

\end{abstract}
\keywords{}

\section{Introduction} 
\label{sec:introduction}

Regardless of its different types of origin, which can be either the collapse of massive star \citep{Paczynski_1986ApJ,Woosley_1993ApJ,woosley2006supernova} or the merger of binary compact stars \citep{eichler1989nucleosynthesis}, a GRB central engine is believed to resemble the same accretion system which consists of a central object, an accretion disk, and a relativistic jet. In particular, if the central object is a black hole (BH), the angular momentum direction of the BH and the accretion disk can differ due to the anisotropic explosions of its progenitor star. As the outer part of the disk has a sufficiently larger angular momentum, it will maintain its direction and drive the BH and the inner part of the accretion disk to precess due to the Lense-Thirring torque and the viscosity of the disk \citep{lense_uber_1918,bardeen1975lense}. The jet launched from the inner region of the disk will follow the rotating black \textcolor{black}{hole} to precess \citep[e.g.,][]{reynoso_precession_2006,liu_jet_2010,lei_frame_2012}. Such precession can naturally cause the change of observer angle (calculated with respect to the moving direction of the ejected material, see \S \ref{subsec:model}). In some cases, when the precession period is shorter than the burst duration, and the jet's opening angle is small enough, precession can affect the observed light curve by introducing periodic-like or missing emissions. Several previous attempts \citep[e.g.,][]{lei_model_2007,liu_jet_2010} have been made to correlate those features with observations.

In light of previous studies, we sought to find additional GRBs that display those characteristics that may be attributed to precession. GRB 220408B, a recent burst co-detected by Fermi \citep{2022GCN.31906....1B}, Konus-Wind \citep{2022GCN.31905....1L}, Astro-Sat \citep{2022GCN.31863....1G}, and GRID (\textit{this work}), quickly caught our attention due to its multiple similar temporal episodes. In this {\it Letter}, we first performed a detailed analysis of GRB 220408B from the perspective of its light curve properties and spectral evolution (\S \ref{sec:dataanalysis}). Motivated by the \textcolor{black}{similarities} of the three episodes in light curve profile, spectral evolution, and spectral lags, we proposed to use a precession-nutation model to explain the observed properties of GRB 220408B (\S \ref{sec:model_fit}). The summary and discussion are presented in \S \ref{sec:result_summary}.

\begin{figure}
 \centering
 \includegraphics[width = 0.47\textwidth]{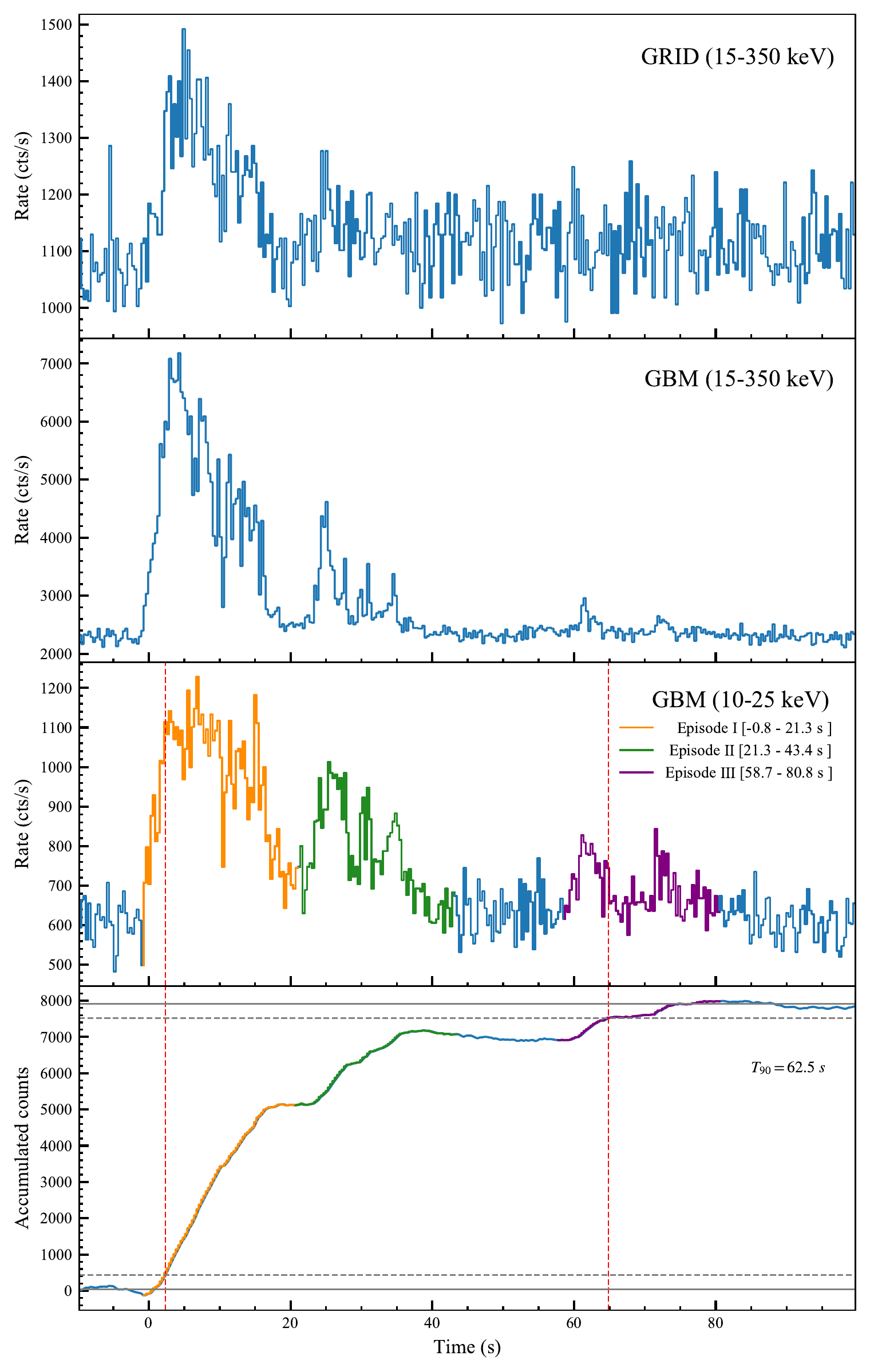}
 \caption{Light curves (top three panels) and accumulated light curve (bottom panel) of GRB 220408B. The bin size is set to 0.325 s for all light curves. The top panel shows the light curve within the energy range from 15 to 350 keV by combining the data from all four GRID detector units. The second panel shows the light curve of NaI detectors n6, n7 and BGO detector b1 of {\it Fermi}/GBM in the energy range of 15-350 keV. The third panel shows the light curve of NaI detectors n6, n7 and BGO detector b1 of {\it Fermi}/GBM in the energy range of 10-25 keV. The bottom panel shows the accumulated counts of the light curve in the third panel. The three episodes are marked in orange, green, and purple, respectively. The gray solid (dashed) lines are drawn at 0\% (5\%) and 100\% (95\%) of the accumulated counts. The vertical red dashed lines in the bottom two panels represent the $T_{90}$ interval.}
 \label{fig:lc&t90}
\end{figure}

\section{Observation and Data Analysis} \label{sec:dataanalysis}

\subsection{The Data}
\label{sec:observation}
GRB 220408B triggered the Gamma-ray Burst Monitor (GBM) aboard the NASA Fermi Gamma-ray Space Telescope \citep{Meegan_2009ApJ, 2009ApJ...697.1071A} at 07:28:04.65 Universal Time on 8 Apr 2022 (hereafter $T_{\rm 0,GBM}$). It is also the third confirmed GRB observed by GRID (short for Gamma-Ray Integrated Detectors), a low-cost project led by students aiming to build an all-sky and full-time CubeSat network to monitor high-energy transient sources, including GRBs, in low Earth orbits \citep{wen2019grid,wen2021compact}. To date, GRID has collected several confirmed GRBs as well as dozens of GRB candidates, of which the first is GRB 210121A \citep{2021ApJ...922..237W}. In this work, we mostly utilize the {\it Fermi}/GBM data in consideration of its wide spectral coverage and high temporal and spectral resolution. As a result of a large separation angle ($\gg$ $60^\circ$) between the pointing direction of the detector and the GRB location, the GRID data of GRB 220408B \textcolor{black}{suffer} from low signal-to-noise ratio\footnote{The signal-to-noise ratio is defined as $S/N=\frac{S-B}{\sigma(B)}$, where $S$ is the count rate in signal region, $B$ is the background rate, $\sigma(B)$ is the variance of $B$.} (S/N) and, therefore, are only displayed in the top panel of Figure \ref{fig:lc&t90} as an illustration.

We retrieved the time-tagged event (TTE) dataset of GRB 220408B from the \emph{Fermi}/GBM public data archive\footnote{\url{https://heasarc.gsfc.nasa.gov/FTP/fermi/data/gbm/daily/}}. Two sodium iodide (NaI) detectors, namely n6 and n7, with the smallest viewing angles with respect to the GRB source direction, were selected for our analysis. Additionally, the brightest bismuth germanium oxide (BGO) detector, b1, was also selected as it extends to a higher energy range. These data were then processed according to the standard procedures described in \cite{2011ApJ...730..141Z} and \cite{2022Natur.612..232Y} to investigate the burst's temporal and spectral properties, as detailed below.

\subsection{The Three-Episode Light Curve} 
\label{subsec:lightcurve}

\begin{figure*}
\centering
\subfigure[]{\includegraphics[width=0.48\textwidth]{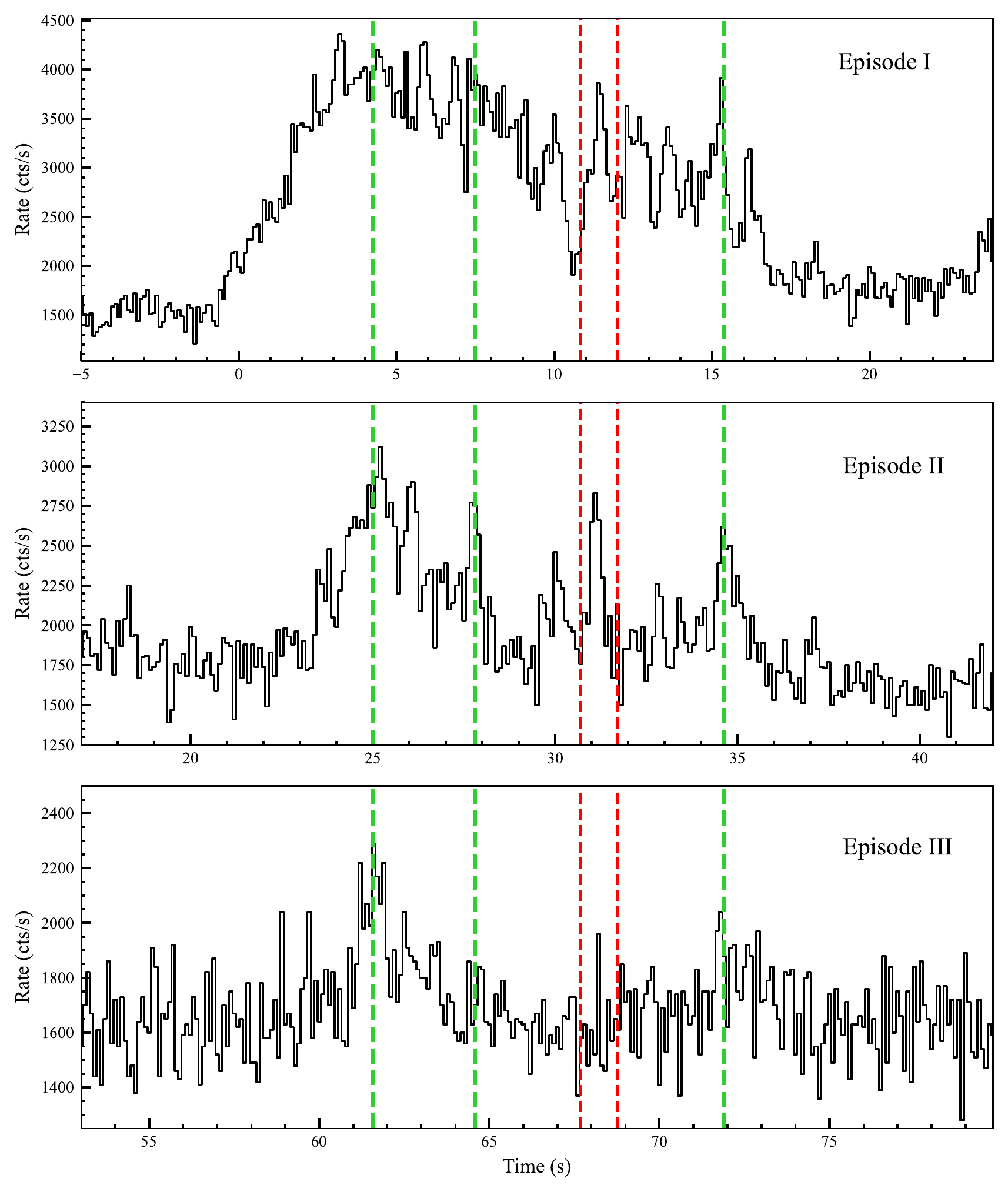}}
\subfigure[]{\includegraphics[width=0.48\textwidth]{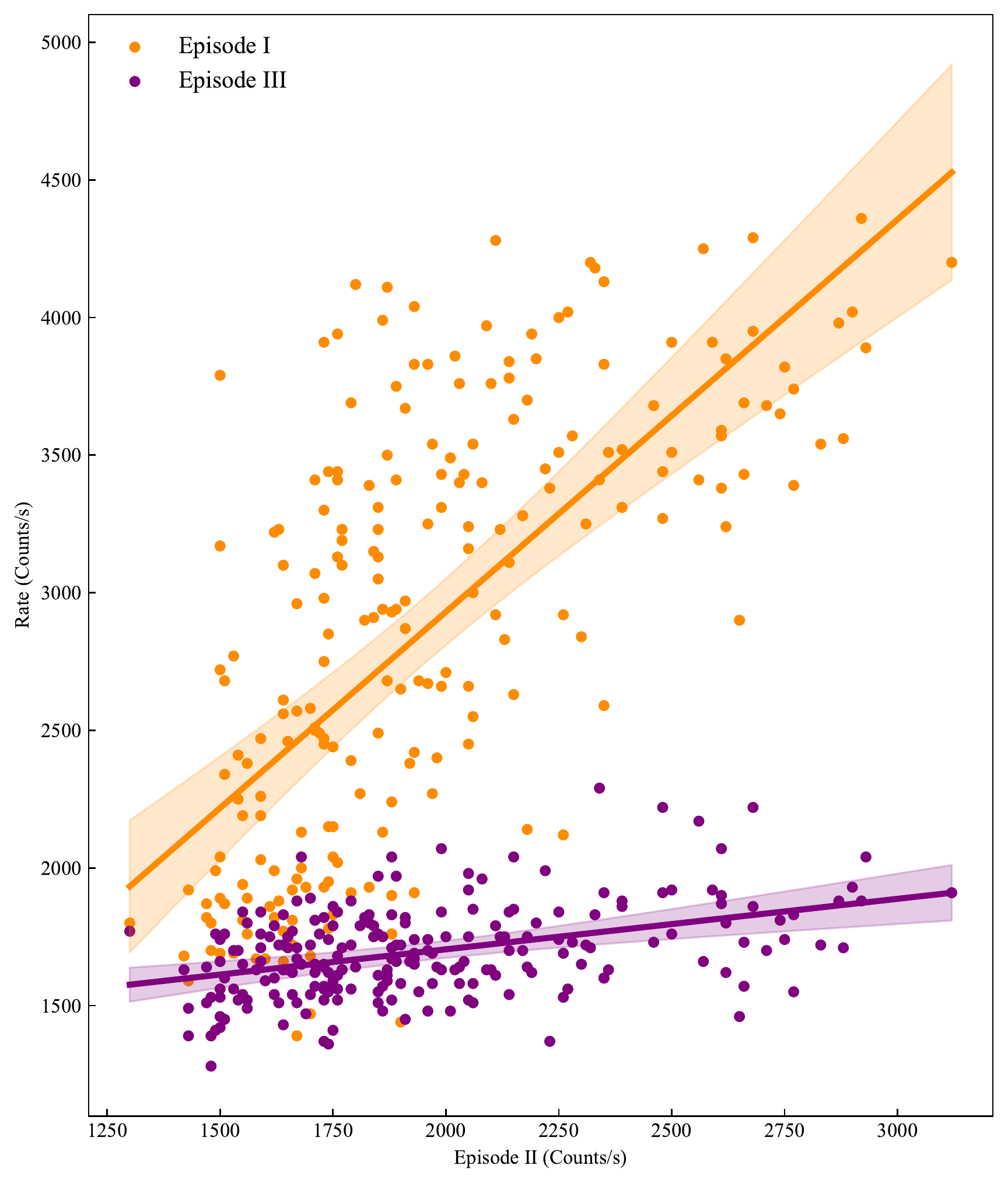}} \\
\caption{(a) Light curve profiles of the three episodes are aligned according to the middle peak positions within red vertical dashed lines. Matched peaks are marked with green vertical dashed lines. (b) The correlation between Episode II and Episode I is delineated in orange, and the correlation between Episode II and Episode III is delineated in purple. The shaded areas represent the corresponding 3$\sigma$ uncertainties.}
\label{fig:lc_comparison}
\end{figure*}

We plot the GBM and GRID light curves together in Figure \ref{fig:lc&t90}, using the same bin size of 0.325 s and the same alignment time, $T_0$, at $T_{\rm 0,GBM}$. While both light curves are extracted from the same 15--350 keV energy range, the $S/N$ of the brightest peak of the GRID light curve is about one-fifth of that of the GBM light curve due to the former's large off-axis angle. Nevertheless, the majority of the significant peaks on both light curves coincide, which reinforces the usefulness of CubeSat detectors for GRB research, even in non-ideal observational conditions.

 GRB 220408B exhibits an overall Fast-Rise-Exponential-Decay \citep[FRED;][]{kocevski_search_2003} profile while retaining a complex substructure characterized by three apparently separated emission episodes. Following the method in \cite{2020ApJ...899..106Y} and \cite{2020ApJ...899...60Y}, the burst duration in the standard energy range of 15-350 keV is calculated as \textcolor{black}{ $T_\textrm{90, 15-350 keV} \sim 30 $} s \citep[see also ][]{2022GCN.31906....1B}, counted from $T_0+1.5$ s to $T_0+31.5$ s. Such a $T_{\rm 90}$ range, however, does not cover the third episode, which starts at around $T_{\rm 0}+55$ s.

 We noticed that the third emission episode becomes particularly significant in low energies, as shown in the third panel \textcolor{black}{of} Figure \ref{fig:lc&t90} and Figure \ref{multi-lc}, indicating a strong spectral evolution across the three episodes. We were thus motivated to recalculate the burst's $T_{90}$ in a lower energy range between 10 keV and 25 keV to be $62.5$ s (see the bottom panel of Figure \ref{fig:lc&t90}), which more accurately conveys the burst time scale and the central engine activities \textcolor{black}{\citep{2014ApJ...787...66Z}}. Such an energy range is also utilized in dividing the burst into three episodes with a visual aid of the pulse structures, as colored in the third panel of Figure \ref{fig:lc&t90}. 

Interestingly, the three emission episodes display striking similarities with each other in terms of duration, pulse structure, and spectral evolution. A more comprehensive analysis of that focus will be conducted in the rest of this section.

\subsection{Similarity in Overall Temporal Profile}
\label{subsec:Similar-lightcurve}

The similarity of the light curve profiles is illustrated in Figure \ref{fig:lc_comparison}(a), where the light curves\footnote{Those light curves are extracted in the energy range of 10-100 keV to improve $S/N$ and binned to 0.1 s to increase the visibility of the detailed structures} of the three episodes are first aligned according to the middle peak positions (red vertical dashed lines). Interestingly, such an alignment automatically results in several other peaks matching along (green vertical dashed lines).

To further quantify the similarity, we calculated the correlation coefficients between any pair of the three light curves in \textcolor{black}{Figure \ref{fig:lc_comparison}(b)}. The Pearson correlation coefficient is \textcolor{black}{0.68} with a p-value of \textcolor{black}{$7.60 \times 10^{-31} $} between Episode I and Episode II, and is \textcolor{black}{0.42} with a p-value of \textcolor{black}{$1.15 \times 10^{-10} $} between Episode II and Episode III. The strong correlations among the three episodes indicate that they may have the same physical origin, which could account for their similar shapes.

\subsection{Similarity in Multi-wavelength Behaviors}

\begin{figure}
\hspace{-0.2cm}
 \includegraphics[width=0.4\textwidth]{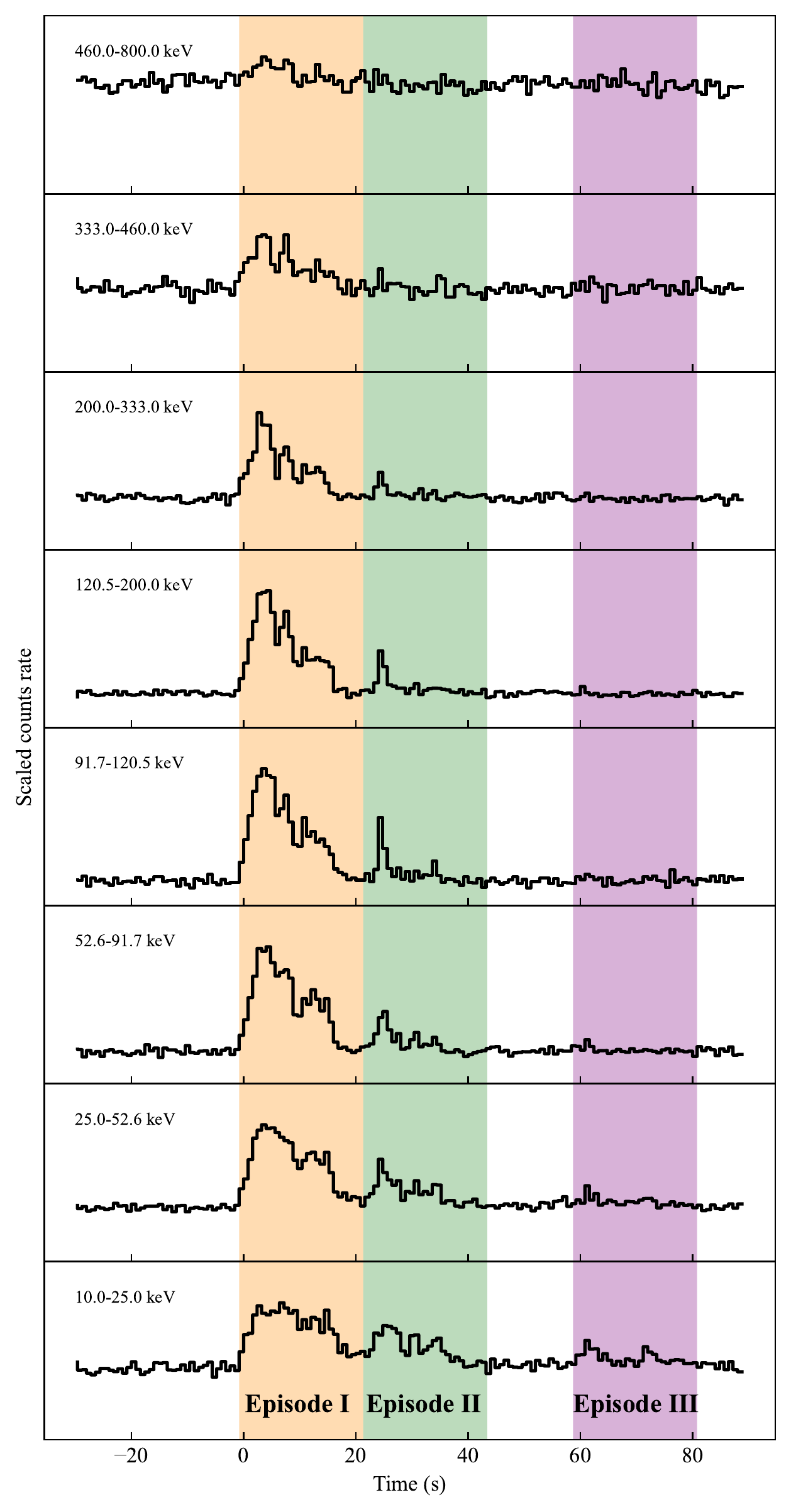}
 \centering
 \caption{The multi-wavelength light curves of GRB 220408B. The orange, green, and purple blocks mark Episode I, Episode II, and Episode III, respectively.}
 \label{multi-lc}
\end{figure}

We then divided the energy range between 10 keV and 800 keV into eight bands and extracted multi-wavelength light curves from {\it Fermi}/GBM NaI detectors n6, n7 and BGO detector b1 using the method described in \cite{2022ApJ...935...79L}. As shown in Figure \ref{multi-lc}, the profiles of the multi-wavelength light curves, including their characteristics, such as the peak time and width of the pulses, clearly evolve in accordance with increasing energy. Such an evolution is commonly observed in GRBs and often measured as spectral lag, which refers to the delay of the arrival time of gamma-ray photons in different energy bands \citep[e.g.,][]{norris2000connection, Yi_2006MNRAS}. Both positive (i.e., higher-energy photons arrive earlier) and negative lags, as well as the positive-to-negative lag transitions \textcolor{black}{\citep[e.g.,][]{2017ApJ...834L..13W,2021ApJ...906....8D,2022ApJ...935...79L}}, have been observed in some GRBs.
\begin{figure}
 \centering
 \includegraphics[width=0.35\textwidth]{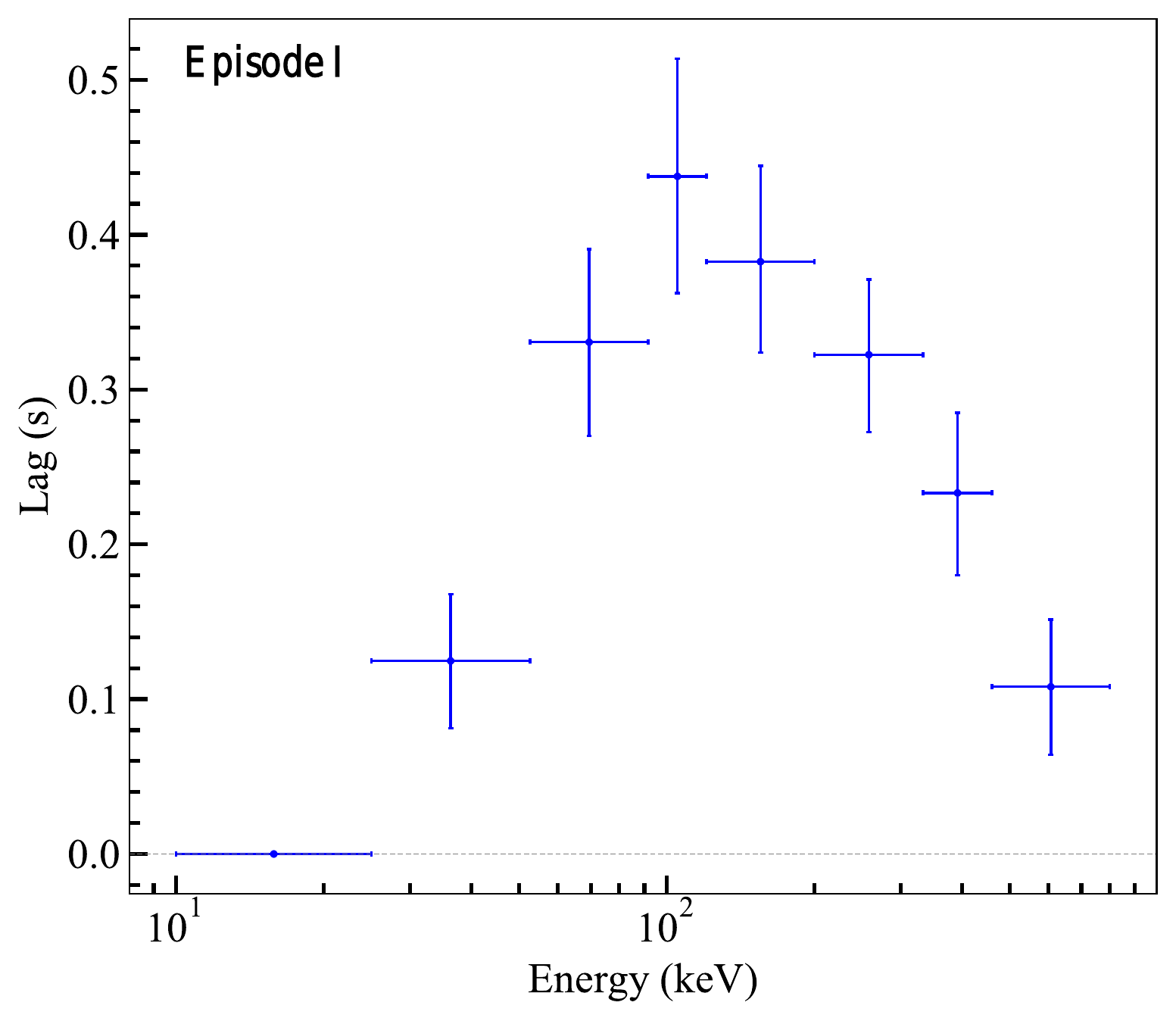}
 \includegraphics[width=0.35\textwidth]{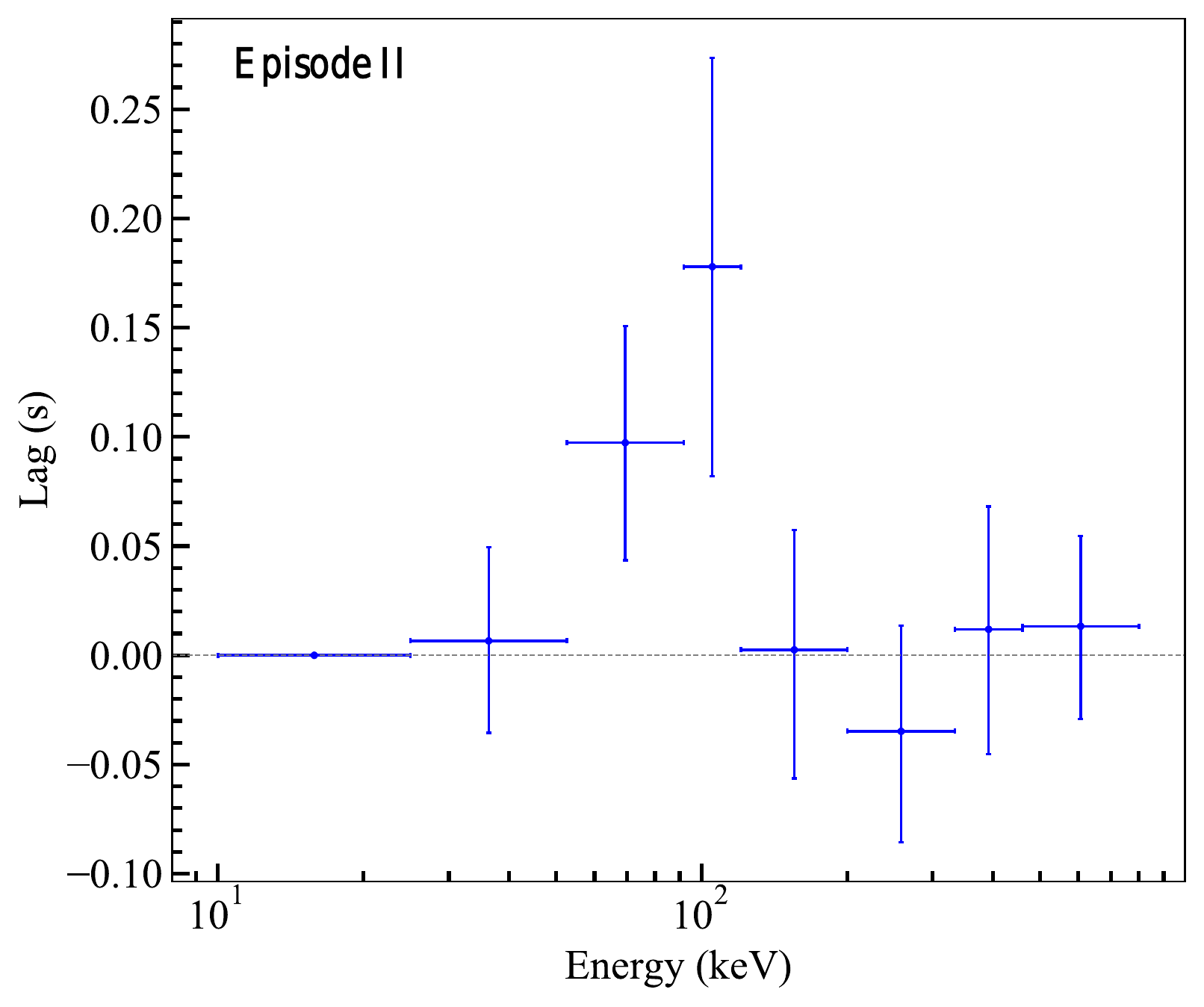}
 \includegraphics[width=0.35\textwidth]{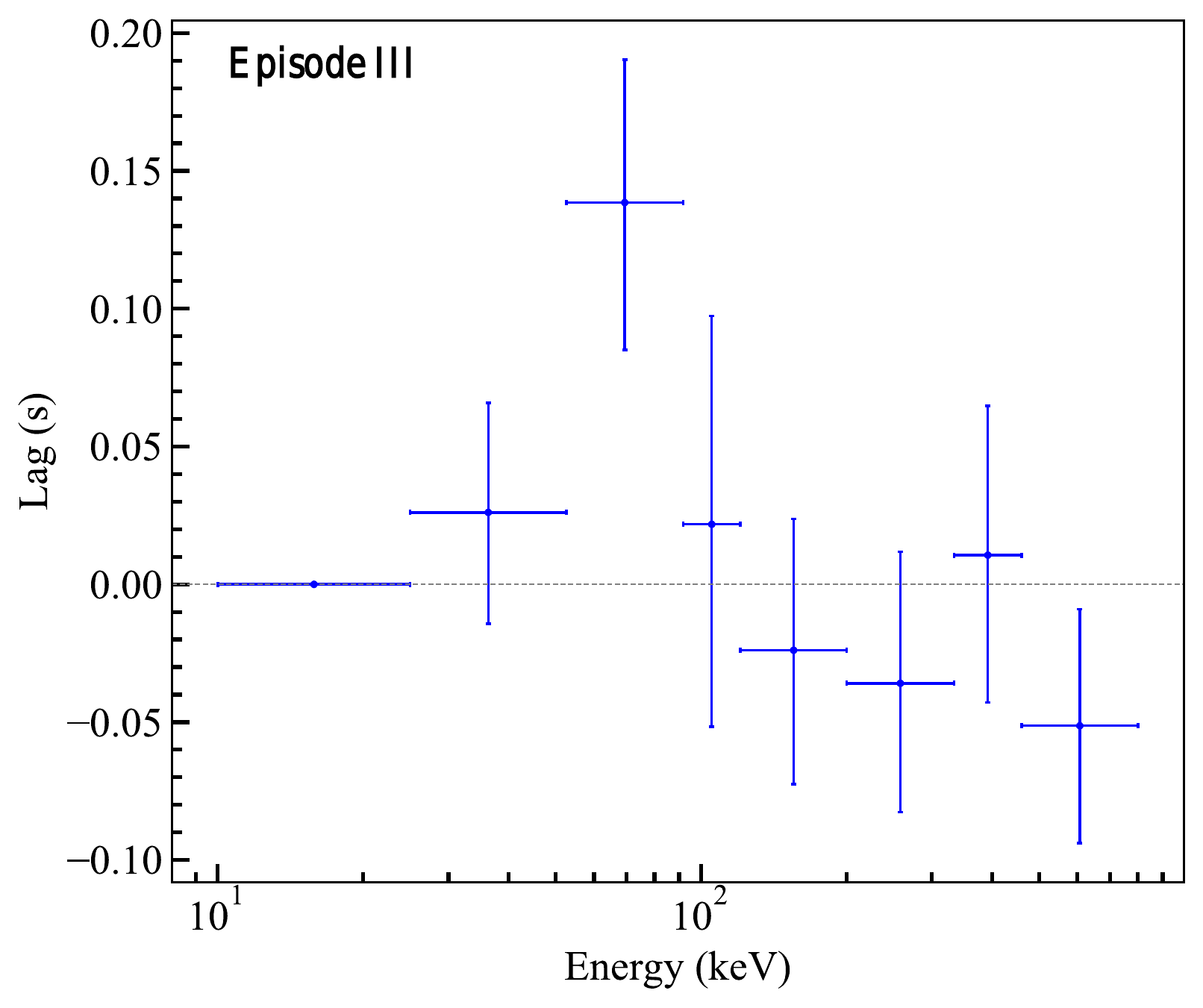}
 \caption{Spectral lags between the lowest energy band (10-25 keV) and any higher energy bands of the three episodes (Episodes I to III from top to bottom). The zero lag is shown with a dashed gray line. The horizontal error bars represent the range of energy bands, and the vertical error bars indicate the 1$\sigma$ uncertainties.}
 \label{fig:lags}
\end{figure}

Following the method described in \cite{Zhang_2012ApJ} and \textcolor{black}{\cite{2022ApJ...935...79L}}, we calculated the energy-dependent lags for all three episodes using the multi-wavelength light \textcolor{black}{curve} pairs in Figure \ref{multi-lc}. The results are shown in Figure \ref{fig:lags}. Interestingly, a positive-to-negative transition feature, with a roughly consistent trend, is observed in the lag-E relations in all three episodes, which further confirms the similarity of the three emission episodes.

\subsection{Similarity in Spectral Evolution} 
\label{sec:spectralfitting}
\begin{figure}
\hspace{-0.2cm}
 \includegraphics[width=0.47\textwidth]{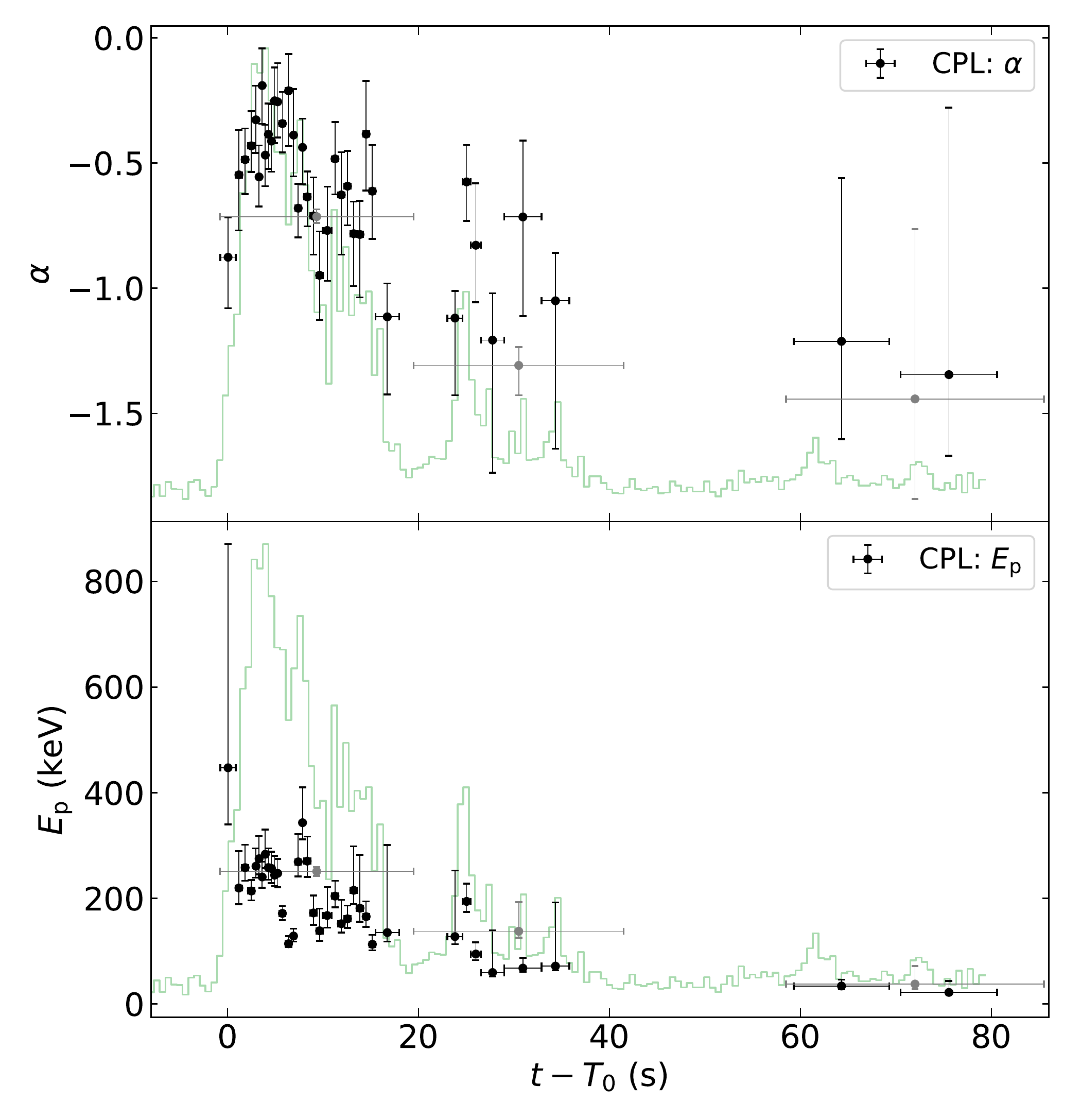}
 \centering
 \caption{The spectral evolution of the CPL model. The horizontal error bars represent the time spans, and the vertical error bars indicate the 1$\sigma$ uncertainties of the best-fit parameters.}
 \label{fig:evolution}
\end{figure}

We performed both time-integrated and time-resolved spectral \textcolor{black}{analyses} over the periods of Episodes I, II, and III, respectively. Our time-resolved spectral analysis seeks to track the spectral evolution in as much detail as possible. To do so, we divided the burst duration
into 37 time-resolved slices (see Figure \ref{fig:evolution} and Table \ref{tab:specfit}), each containing sufficient photon counts \textcolor{black}{\citep[i.e., 20 counts per spectral bin;][]{Zhang_2018NatAs} } to ensure statistical validity. Within each slice, we extracted the count spectra of GRB 220408B from \textcolor{black}{{\it Fermi}}/GBM NaI detectors n6, n7 and BGO detector b1 following the procedures described in \cite{2011ApJ...730..141Z,2020ApJ...899...60Y,2020ApJ...899..106Y,2022Natur.612..232Y} and \cite{Zou_2021ApJ}. Corresponding background spectra are acquired by applying the baseline method\textcolor{black}{ } \citep{2020ApJ...899..106Y,Zou_2021ApJ} to the time interval from $T_0-98$ s to $T_0+134$ s for each energy channel. The response matrices of the detectors are generated using the GBM Response Generator\footnote{\url{https://fermi.gsfc.nasa.gov/ssc/data/analysis/rmfit/gbmrsp-2.0.10.tar.bz2}}.

\begin{deluxetable*}{ccclcc|ccclcc}
\tablecaption{The Spectral Fitting Results and Corresponding Energy Flux of GRB 220408B \label{tab:specfit}}
\tabletypesize{\tiny}
\tablehead{
\colhead{t1 (s)} & \colhead{t2 (s)} & \colhead{$\alpha$} & \colhead{$E_{\rm p}$ (keV)} & \colhead{flux ($\rm erg\,cm^{-2}\,s^{-1}$)} & \colhead{pgstat/dof} & \colhead{t1 (s)} & \colhead{t2 (s)} & \colhead{$\alpha$} & \colhead{$E_{\rm p}$ (keV)} & \colhead{flux ($\rm erg\,cm^{-2}\,s^{-1}$)} & \colhead{pgstat/dof}
}
\startdata
{-}0.80&19.50&${-0.71}_{-0.03}^{+0.03}$&${251.11}_{-8.66}^{+8.25}$&${1.54}_{-0.04}^{+0.03}\times10^{-6}$&410.2/359& 8.02&8.67&${-0.64}_{-0.12}^{+0.10}$&${270.49}_{-30.14}^{+46.34}$&${2.39}_{-0.18}^{+0.26}\times10^{-6}$&221.6/358 \\
19.50&41.50&${-1.31}_{-0.12}^{+0.07}$&${137.72}_{-12.17}^{+55.23}$&${2.02}_{-0.09}^{+0.34}\times10^{-7}$&239.2/359& 8.67&9.32&${-0.71}_{-0.16}^{+0.15}$&${172.39}_{-22.53}^{+33.03}$&${1.29}_{-0.11}^{+0.14}\times10^{-6}$&233.1/358 \\
58.50&85.50&${-1.44}_{-0.40}^{+0.68}$&${37.81}_{-9.60}^{+34.02}$&${1.48}_{-0.48}^{+2.56}\times10^{-8}$&178.0/359& 9.32&9.97&${-0.95}_{-0.18}^{+0.18}$&${138.58}_{-18.99}^{+41.78}$&${8.07}_{-0.78}^{+1.17}\times10^{-7}$&226.3/358 \\
\cline{1-6} 
{-}0.80&0.87&${-0.88}_{-0.20}^{+0.16}$&${447.15}_{-107.36}^{+423.33}$&${8.32}_{-1.50}^{+3.97}\times 10^{-7}$&199.0/358& 9.97&10.92&${-0.77}_{-0.20}^{+0.18}$&${167.52}_{-23.04}^{+54.17}$&${7.23}_{-0.73}^{+1.20}\times10^{-7}$&220.0/358 \\
0.87&1.52&${-0.55}_{-0.22}^{+0.18}$&${219.60}_{-30.55}^{+69.45}$&${1.08}_{-0.12}^{+0.21}\times10^{-6}$&246.3/358& 10.92&11.60&${-0.48}_{-0.14}^{+0.15}$&${204.27}_{-21.13}^{+28.86}$&${1.63}_{-0.13}^{+0.14}\times10^{-6}$&231.0/358 \\
1.52&2.17&${-0.49}_{-0.14}^{+0.12}$&${258.16}_{-24.73}^{+43.52}$&${2.12}_{-0.15}^{+0.24}\times10^{-6}$&230.9/358& 11.60&12.25&${-0.63}_{-0.24}^{+0.17}$&${151.99}_{-16.79}^{+45.27}$&${9.11}_{-0.77}^{+1.54}\times10^{-7}$&228.2/358\\
2.17&2.82&${-0.43}_{-0.10}^{+0.14}$&${213.95}_{-17.85}^{+20.71}$&${2.18}_{-0.14}^{+0.14}\times10^{-6}$&255.9/358& 12.25&12.90&${-0.59}_{-0.16}^{+0.14}$&${161.25}_{-17.21}^{+25.00}$&${1.27}_{-0.10}^{+0.12}\times10^{-6}$&240.0/358\\
2.82&3.15&${-0.33}_{-0.13}^{+0.14}$&${260.95}_{-22.00}^{+33.64}$&${3.15}_{-0.21}^{+0.29}\times10^{-6}$&251.0/358& 12.90&13.55&${-0.78}_{-0.21}^{+0.13}$&${215.28}_{-25.75}^{+83.26}$&${1.13}_{-0.10}^{+0.23}\times10^{-6}$&255.8/358\\
3.15&3.47&${-0.56}_{-0.12}^{+0.13}$&${275.12}_{-29.90}^{+42.90}$&${3.45}_{-0.29}^{+0.34}\times10^{-6}$&253.2/358& 13.55&14.20&${-0.79}_{-0.25}^{+0.13}$&${181.39}_{-25.49}^{+101.17}$&${1.03}_{-0.10}^{+0.29}\times10^{-6}$&240.2/358\\
3.47&3.80&${-0.19}_{-0.15}^{+0.15}$&${240.45}_{-20.51}^{+28.94}$&${3.10}_{-0.23}^{+0.28}\times10^{-6}$&249.3/358& 14.20&14.85&${-0.38}_{-0.23}^{+0.21}$&${165.44}_{-19.43}^{+29.04}$&${9.70}_{-0.88}^{+1.19}\times10^{-7}$&223.9/358\\
3.80&4.12&${-0.47}_{-0.12}^{+0.12}$&${283.50}_{-26.42}^{+46.84}$&${3.48}_{-0.24}^{+0.38}\times10^{-6}$&239.7/358& 14.85&15.50&${-0.61}_{-0.19}^{+0.19}$&${113.02}_{-11.58}^{+17.89}$&${9.06}_{-0.67}^{+0.90}\times10^{-7}$&207.8/358\\
4.12&4.45&${-0.39}_{-0.14}^{+0.12}$&${258.24}_{-23.10}^{+36.51}$&${3.24}_{-0.20}^{+0.33}\times10^{-6}$&230.1/358& 15.50&17.97&${-1.11}_{-0.31}^{+0.13}$&${135.27}_{-16.88}^{+165.56}$&${3.55}_{-0.26}^{+1.55}\times10^{-7}$&245.7/358\\
4.45&4.77&${-0.41}_{-0.12}^{+0.15}$&${256.58}_{-27.64}^{+31.79}$&${3.30}_{-0.29}^{+0.28}\times10^{-6}$&242.0/358& 23.02&24.62&${-1.12}_{-0.31}^{+0.11}$&${127.66}_{-14.25}^{+125.11}$&${3.25}_{-0.64}^{+1.68}\times10^{-7}$&275.8/358\\
4.77&5.10&${-0.25}_{-0.17}^{+0.13}$&${243.95}_{-21.16}^{+36.75}$&${2.97}_{-0.21}^{+0.32}\times10^{-6}$&242.7/358& 
24.62&25.47&${-0.58}_{-0.16}^{+0.15}$&${194.24}_{-19.98}^{+33.75}$&${1.13}_{-0.09}^{+0.13}\times10^{-6}$&252.0/358\\
5.10&5.42&${-0.26}_{-0.14}^{+0.16}$&${247.39}_{-26.54}^{+27.43}$&${3.12}_{-0.28}^{+0.23}\times10^{-6}$&231.8/358& 25.47&26.55&${-0.83}_{-0.23}^{+0.25}$&${94.63}_{-11.42}^{+21.93}$&${4.31}_{-0.38}^{+0.54}\times10^{-7}$&233.0/358\\
5.42&6.07&${-0.34}_{-0.12}^{+0.13}$&${171.57}_{-12.94}^{+14.04}$&${1.98}_{-0.12}^{+0.12}\times10^{-6}$&230.2/358& 26.55&28.97&${-1.21}_{-0.53}^{+0.19}$&${59.63}_{-7.61}^{+79.50}$&${1.92}_{-0.14}^{+0.94}\times10^{-7}$&205.8/358\\
6.07&6.72&${-0.21}_{-0.22}^{+0.15}$&${114.16}_{-6.60}^{+14.13}$&${1.15}_{-0.06}^{+0.09}\times10^{-6}$&219.0/358& 28.97&32.90&${-0.72}_{-0.40}^{+0.31}$&${67.98}_{-7.07}^{+19.65}$&${1.48}_{-0.13}^{+0.24}\times10^{-7}$&241.8/358\\
6.72&7.12&${-0.39}_{-0.17}^{+0.18}$&${128.75}_{-10.63}^{+13.90}$&${1.42}_{-0.10}^{+0.10}\times10^{-6}$&241.3/358& 32.90&35.80&${-1.05}_{-0.59}^{+0.19}$&${71.80}_{-8.31}^{+120.63}$&${1.69}_{-0.14}^{+0.87}\times10^{-7}$&201.4/358\\
7.12&7.70&${-0.68}_{-0.12}^{+0.10}$&${269.13}_{-27.57}^{+52.14}$&${2.43}_{-0.17}^{+0.29}\times10^{-6}$&231.6/358& 59.30&69.30&${-1.21}_{-0.39}^{+0.65}$&${34.11}_{-6.72}^{+12.02}$&${3.20}_{-0.80}^{+0.93}\times10^{-8}$&253.2/359\\
7.70&8.02&${-0.44}_{-0.15}^{+0.11}$&${343.19}_{-31.41}^{+67.14}$&${3.81}_{-0.28}^{+0.52}\times10^{-6}$&240.2/358& 70.50&80.60&${-1.35}_{-0.32}^{+1.07}$& ${22.06}_{-4.38}^{+21.50}$&${1.55}_{-0.68}^{+0.97}\times10^{-8}$&227.3/359\\
 \enddata
\end{deluxetable*}

\begin{figure}
 \centering
 \includegraphics[angle=0,width=0.47 \textwidth]{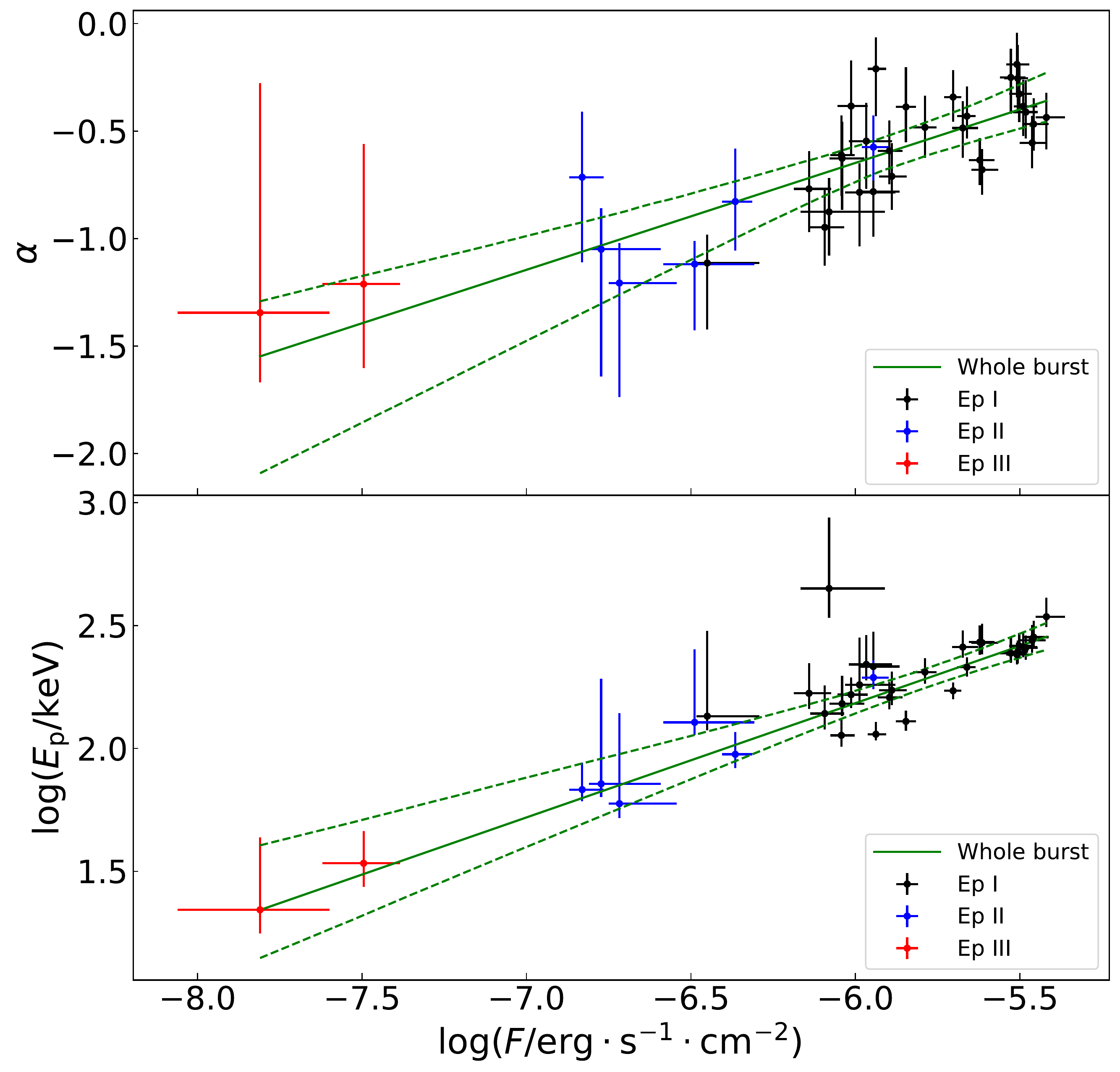}
 \caption{The linear fits to $\log E_{\rm p} - \log F$ and $\alpha - \log F$ relations. The solid green lines and dashed green lines show the best-fit relations and the corresponding 3$\sigma$ error bands of the whole burst, respectively. The black, blue, and red points are the parameters of Episode I, Episode II, and Episode III, respectively. The error bars indicate the corresponding 1$\sigma$ uncertainties of the parameters.
 }
 \label{fig:para_relation}
\end{figure}
For each slice, as well as each episode, we performed a spectral fit using the Monte Carlo fitting tool \emph{MySpecFit} \citep{2022Natur.612..232Y}. All the spectra are fitted by a cutoff power-law (CPL) model formulated as \citep{2016A&A...588A.135Y}
\begin{equation}
 N(E)=A{E}^{\alpha}e^{\frac{-(\alpha+2)E}{E_{\rm p}}},
\end{equation}
where $\alpha$, $A$, and $E_{\rm p}$ are the photon index, normalization coefficient, and peak energy, respectively. 

The results of our spectral fitting are presented in Table \ref{tab:specfit}. Based on the ratio of Profile Gaussian likelihood to the degree of freedom \citep[PGSTAT/dof;][]{1996ASPC..101...17A} %the Bayesian information criterion \citep[BIC;][]{Schwarz_1978AnSta} 
statistics, our results indicate that the CPL model can adequately fit the spectra of the three episodes and time-resolved slices. Using the best-fit parameters of the CPL model, we plot the spectral evolution of GRB 220408B in Figure \ref{fig:evolution}, along with the total light curve summing up the GBM detectors n6, n7, and b1 \textcolor{black}{between 10 and 1,000 keV}. Both $\alpha$ and $E_{\rm p}$ exhibit strong spectral evolution, roughly consistent with the tracking behaviors as observed in other GRBs \citep[e.g.,][]{Lu_2012ApJ}.

Further analysis of the spectral evolution is carried out on an episode-by-episode basis. In Figure \ref{fig:para_relation}, we plot $\alpha$ and $E_{\rm p}$ as a function of the \textcolor{black}{flux}, $F$, \textcolor{black}{in 10-10,000 keV} of each time slice for the three episodes. One can see those observable pairs are strongly correlated and consistently follow the same tracks of \textcolor{black}{ log$E_{\rm p} =0.46^{+0.03}_{-0.03}{\rm log} F +4.95^{+0.17}_{-0.16}$ } and \textcolor{black}{ $\alpha=0.56^{+0.07}_{-0.08}{\rm log} F +2.74^{+0.43}_{-0.47}$}, respectively. Such consistency suggests that the spectral evolution patterns of the three episodes are similar, despite the fact that the global spectra of the burst undergo a hard-to-soft transition, which is typically attributed to GRB central engine characteristics \citep[e.g.,][]{Zhang_2018NatAs}.

\begin{figure}
 \centering
 \includegraphics[angle=0,width=0.4\textwidth,trim=20 100 20 100,clip]{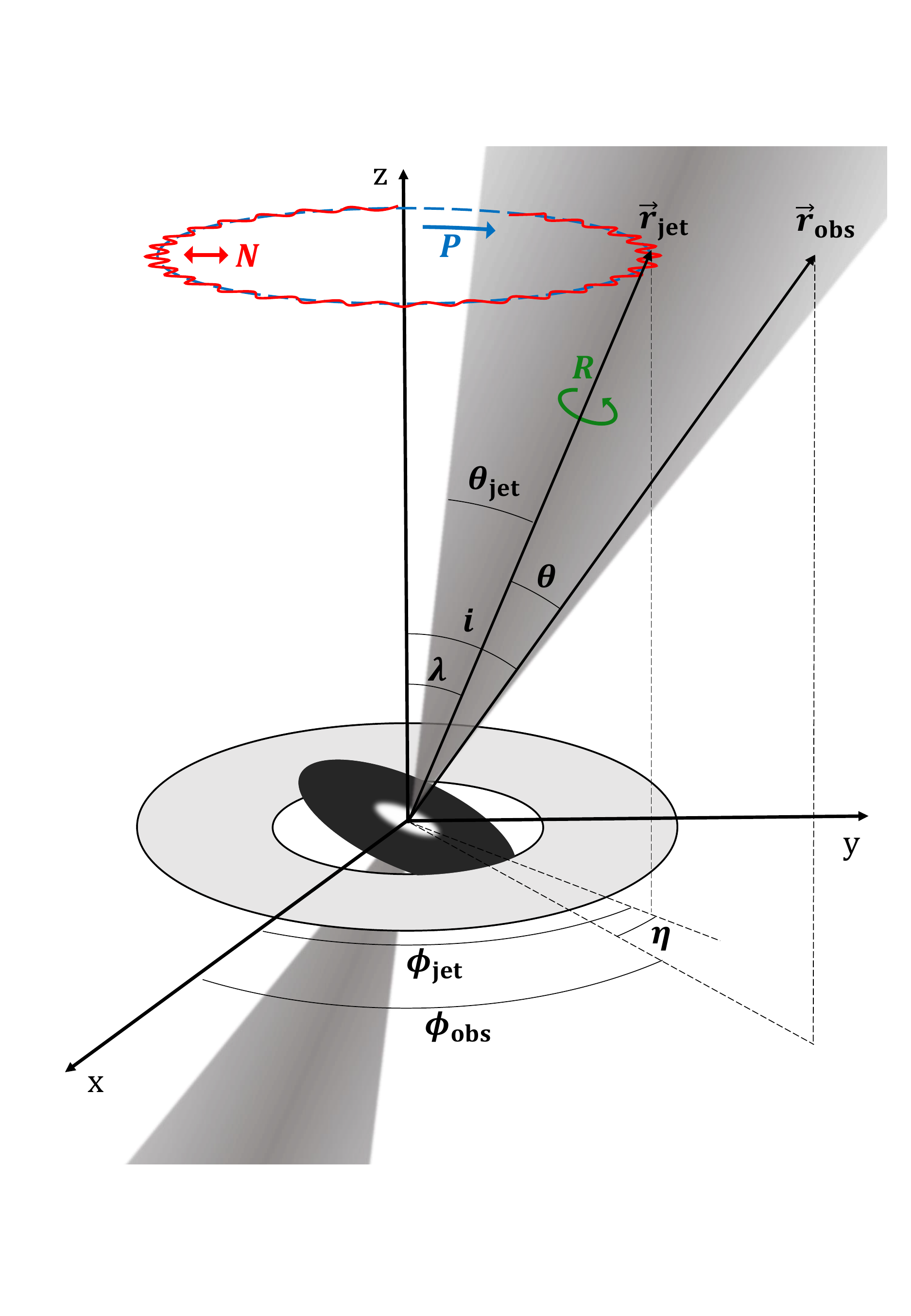}
 \caption{The schematic sketch of the precession-nutation jet model.
 \begin{CJK}{UTF8}{gbsn}
\end{CJK}
 }
 \label{fig:jet_sch}
\end{figure}

\begin{table*}
\centering
\caption{The Best-Fit Parameters of Linear Models for $\alpha-\log F$, $\log E_{\rm p}-\log F$, and $\log E_{\rm p}-\alpha$ Correlations}
\label{tab:alpha_Ep_flux}
\begin{tabular}{cccccccccc}
\hline
\hline
\multirow{2}{*}{Line model}&\multicolumn{2}{c}{Episode I}&&\multicolumn{2}{c}{Episode II}&&\multicolumn{2}{c}{Whole burst}\\
\cline{2-3}
\cline{5-6}
\cline{8-9}
&$k$&$b$&&$k$&$b$&&$k$&$b$\\
\hline
$\alpha=k{\rm log}F+b$&$0.70_{-0.09}^{+0.08}$&$3.48_{-0.53}^{+0.47}$&&$0.67_{-0.20}^{+0.24}$&$3.39_{-1.22}^{+1.54}$&&$0.56_{-0.08}^{+0.07}$&$2.74_{-0.47}^{+0.43}$\\ 
${\rm log}E_{\rm p}=k{\rm log}F+b$&$0.47_{-0.06}^{+0.05}$&$4.99_{-0.32}^{+0.31}$&&$0.55_{-0.10}^{+0.11}$&$5.57_{-0.62}^{+0.68}$&&$0.46_{-0.03}^{+0.03}$&$4.95_{-0.16}^{+0.17}$\\ 
${\rm log}E_{\rm p}=k\alpha+b$&$0.06_{-0.19}^{+0.25}$&$2.33_{-0.11}^{+0.14}$&&$0.30_{-0.43}^{+0.39}$&$2.30_{-0.41}^{+0.39}$&&$0.49_{-0.09}^{+0.09}$&$2.56_{-0.06}^{+0.06}$&\\ 
\hline
\end{tabular}
\end{table*}

\section{Model and Fit}
\label{sec:model_fit}

The similarities of the three emission episodes, as well as the overall FRED shape profile, appear to point toward a uniform origin that produces the observed gamma-ray emissions in a repeatable manner. A natural explanation for such features is that the GRB jet may precess while propagating outward from the central engine \citep{zwart1999can,lei_model_2007,liu_jet_2010}. In this section, we further test this hypothesis by quantitatively fitting the observed data with a precession model. In addition to the precession itself, our toy model also considers that the nutation \citep{zwart1999can} of the jet can contribute to the substructure of the light curves.

\subsection{The Precession-Nutation Model}
\label{subsec:model}

Considering the precession and nutation of a GRB jet, the observer angle, $\theta$, defined as the angle between the jet propagating direction and line of sight (LOS), varies as a function of time. A GRB can be significantly observed only when $\theta$ is less than the jet's half-opening angle. The periodic change of $\theta$ may cause the jet to sweep across the LOS intermittently, which, when taking into account the intrinsic emission profiles together, can lead to complex shapes of GRB light curves, sometimes with repeating \citep[][]{zwart1999can, lei_model_2007} and emission-missing \textcolor{black}{\citep[][]{2022ApJ...931L...2W}} features.

Our model is illustrated in Figure \ref{fig:jet_sch}. A GRB jet with a half-opening angle, $\theta_{\rm jet}$, is propagating along its direction of \textcolor{black}{$\hat{r}_{\rm jet}$}. An observer resides within the $\theta_{\rm jet}$-cone with an off-axis angle, $\theta$, with respect to \textcolor{black}{$\hat{r}_{\rm jet}$}. The jet precesses with an angular velocity of $\omega_{\rm pre}$ along the z-axis while its rotating axis is nutating with an angular velocity of $\omega_{\rm nu}$. The x-y plane is set accordingly so the Cartesian coordinate is centered at the GRB central engine. We also assumed the intrinsic emission from the jet is shaped as a FRED function \citep{kocevski_search_2003}, namely, 
\begin{equation}
\label{equ:FRED}
 F_{\rm o}(t) = F_{\rm m}\left(\frac{t}{t_{\rm m}}\right)^r\left[\frac{d}{d+r}+\frac{r}{d+r}\left(\frac{t}{t_{\rm m}}\right)^{r+1}\right]^{-\frac{r+d}{r+1}},
\end{equation}
where $t$ is measured in \textcolor{black}{the laboratory frame} (the jet's local frame), $t_{\rm m}$ is the time when the flux reaches the peak, $F_{\rm m}$ is the peak flux, $r$ and $d$ are the power-law exponents for the rise and decay, respectively.

We then derived the observed flux of our model based on the above configuration. The direct effect brought by precession and nutation is the change of $\theta$ as a function of time, which can be calculated as
\begin{equation}
\label{equ:openangle}
\begin{aligned}
\theta(t) &= \arccos(\hat{r}_{\rm jet}(t) \cdot \hat{r}_{\rm obs} )\\
 &=\arccos[\cos(\lambda(t)) \cos i+\sin(\lambda(t)) \sin i \cos(\eta(t))], 
\end{aligned}
\end{equation}
where $\lambda(t)$ is the angle between $\hat{r}_{\rm jet}(t)$ and z-axis, $i$ is the angle between $\hat{r}_{\rm obs}$ and z-axis. $\eta (t)$ is defined as 
\begin{equation}
 \eta(t) = \phi_{\rm jet}(t)-\phi_{\rm obs},
\end{equation}
where $\phi_{\rm jet}$ and $\phi_{\rm obs}$ represent the separation angles of directions of the jet ($\hat{r}_{\rm jet}$) and observer ($\hat{r}_{\rm obs}$) to the x-axis, respectively.

According to the kinematical description of the angular evolution of the jet resulting from the precession and nutation \citep{zwart1999can}, $\lambda$ and $\phi_{\rm jet}$ can be expressed as
\begin{equation}
 \lambda(t) = \lambda_0 + \frac{\omega_{\rm pre}}{\omega_{\rm nu}}\tan\lambda_0\cos(\omega_{\rm nu}t) ,
\end{equation}
\begin{equation}
\label{equ:phi_jet}
 \phi_{\rm jet}(t) = \phi_{\rm jet,0} + \omega_{\rm pre}t + \frac{\omega_{\rm pre}}{\omega_{\rm nu}}\sin(\omega_{\rm nu}t).
\end{equation}

The increasing time intervals between the three episodes may be due to the slowing down of jet precession, which is a natural outcome of energy dissipation. We thus assumed a power-law decay of the precession angular velocity as

\begin{equation}
\label{equ:omega_decay}
 \omega_{\rm pre}(t) = \omega_{\rm pre,0}\left(\frac{t-t_0}{t_{\rm C}}\right)^{-\xi},
\end{equation}
where $\xi$ is the decay index, $t_0$ is the offset time when the jet begins to precess, $t_{\rm C}$ is a characteristic time scale.

We assumed a conical jet with a half-opening angle of $\theta_{\rm jet}$ and no moving material outside the cone. According to the derivation in \cite{2016MNRAS.461.3607S}, the observed GRB flux $M(t,\theta(t))$ at viewing angle $\theta(t)$ in the laboratory frame can be described by
\begin{equation}
\label{equ:fluxratio}
 \begin{aligned}
 M(t,\theta(t))& \\
 =& F_{\rm o}(t,\theta=0) \\
 \times & 
		\begin{cases}
1 ,&\theta(t) \leq \theta_{\rm jet}^{*},\\
\displaystyle 1-\frac{\Gamma(\theta(t)-\theta^{*}_{\rm jet})}{2}, & \theta_{\rm jet}^{*}<\theta(t)\leq \theta_{\rm jet},\\
\displaystyle\frac{1}{2}\left[\frac{D}{(1+\beta)\Gamma}\right]^{4-\sqrt{2}\theta_{\rm jet}^{1/3}} , &\theta(t)>\theta_{\rm jet},
\end{cases}
\end{aligned}
\end{equation}
where $F_{\rm o}(t,\theta=0)$ is the observed intrinsic flux when the LOS is centered on the jet axis, $\theta_{\rm jet}^{*} = \theta_{\rm jet} - \frac{1}{\Gamma}$, $\Gamma = (1-\beta^2)^{-1/2}$ is the Lorentz factor, $\beta$ is the dimensionless radial velocity of the jet (i.e., $\beta=v_{\rm jet}/c$, $c$ is the speed of light, $v_{\rm jet}$ is the speed of the jet), and $D$ is the Doppler factor defined as $D = \frac{1}{\Gamma[1-\beta \cos (\theta(t) - \theta_{\rm jet})]}$. For GRBs, the Lorentz factor, $\Gamma$, is typically a few hundred. In this work, we fix $\Gamma$ to be $\Gamma=300$ in consideration that the bulk Lorentz factor does not significantly vary during the prompt emission phase. We also verified that our fitting result is not significantly affected by different values of $\Gamma$.

Furthermore, one needs to convert the time in the laboratory frame to the observer frame by \citep[e.g.,][]{2018pgrb.book.....Z} 
\begin{equation}
\label{equ:time_convert}
 t_{\rm obs} = \frac{1-\beta {\rm cos}\theta(t)}{1-\beta}t_{\rm }-\Delta t,
\end{equation}
where $\Delta t$ is a parameter for adjusting the time offset of the model light curve.

Finally, the observed flux can be calculated by substituting Eqs. \ref{equ:FRED}-\ref{equ:omega_decay} and Eq. \ref{equ:time_convert} to Eq. \ref{equ:fluxratio}, which can be written in form of

\begin{equation}
\label{equ:model2}
\begin{aligned}
 M(t_{\rm obs}) &= M(t_{\rm obs},\mathcal{P}),
 %& =F_{\rm o}(t)[(1-\beta^2)^{1/2}[1-\beta \cos \theta(t)]^{-1}]^{3 + \alpha}
 \end{aligned}
\end{equation}
where $\mathcal{P}$ represents the parameter set as \textcolor{black}{$\mathcal{P} \equiv \{\lambda_0,\,i,\,\eta_{\rm 0},\,\omega_{\rm pre,0},\, \omega_{\rm nu},\,\xi,\,t_{\rm 0},\,t_{\rm C},\,t_{\rm m},\,r,\,d,\, F_{\rm m},\, \Delta t,\, \theta_{\rm jet}, \,\Gamma \}$}, in which $\eta_{\rm 0}$ is defined as $\eta_{\rm 0} = \phi_{\rm jet,0}-\phi_{\rm obs} $.

\begin{figure*}
 \centering
 \includegraphics[width = 0.94\textwidth]{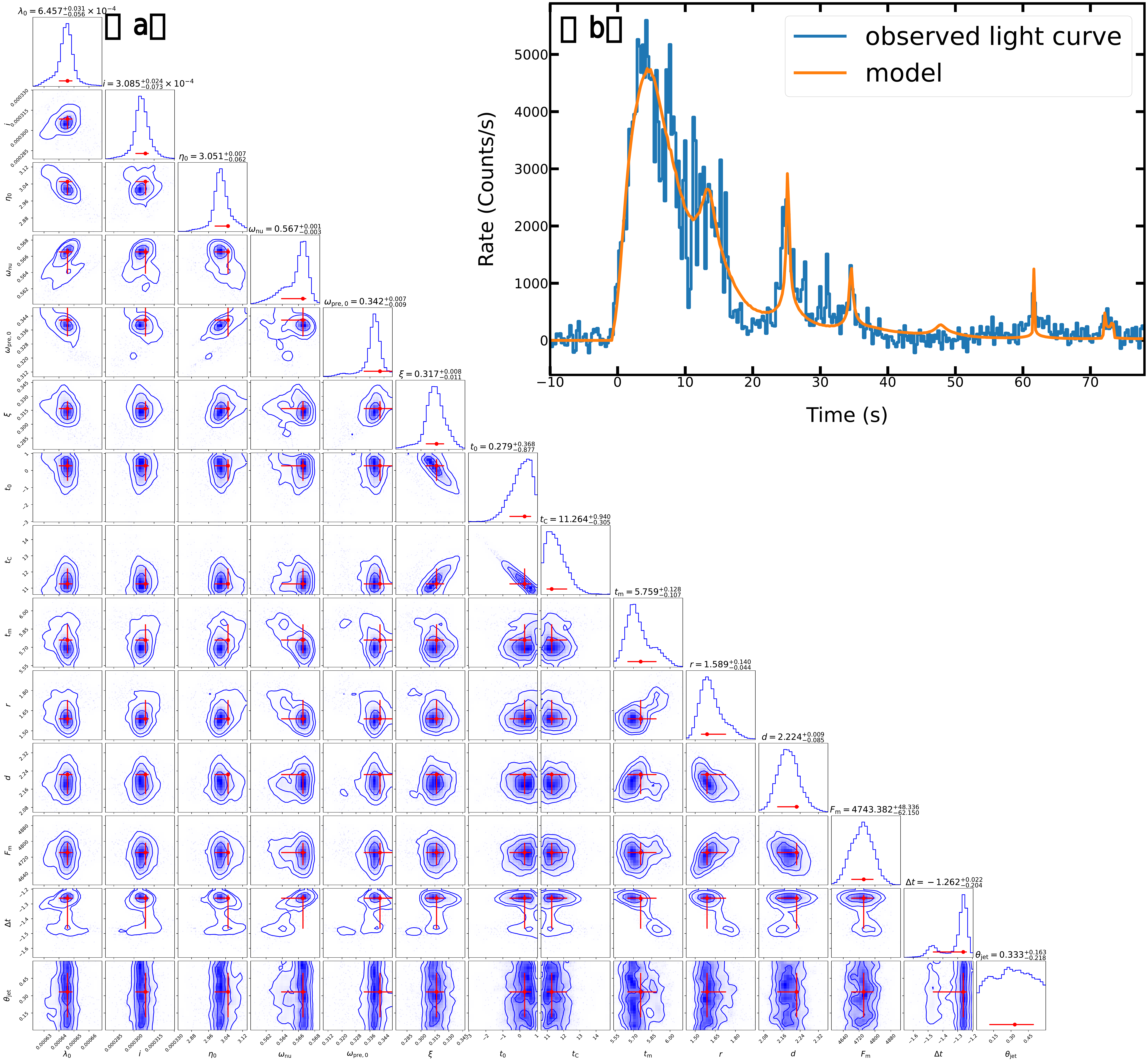}
 \caption{(a) Corner plot of the posterior probability distributions of the parameters. The red error bars represent the 1$\sigma$ uncertainties. The contours in the 2D histograms represent 1$\sigma$, 1.5$\sigma$, and 2$\sigma$ uncertainties, respectively. (b) Light curves of GRB 220408B and the best-fit precession-nutation model.}
 \label{fig:precess_model_fit}
\end{figure*}

\subsection{The Fit}
\label{subsec:fit}

The next step is to fit our model (Eq. \ref{equ:model2}) to the observed light curve. With a fixed parameter of $\Gamma=300$, the free parameter set $P$ of our model consists of the following fourteen items: 

\begin{itemize}
 \item The jet's half-opening angle $\theta_{\rm jet}$. According to \citet{2015ApJ...799....3R}, the maximum of $\theta_{\rm jet}$ is smaller than 0.5 radians. Thus the prior of $\theta_{\rm jet}$ is set as a uniform distribution between 0 and 0.5 radians.
 \item The initial precession angle $\lambda_0$. The prior of $\lambda_0$ is set as a uniform distribution between 0 and $\pi/2$ radians.
 \item The observer's polar angle $i$. The prior of $i$ is set as a uniform distribution between 0 and $\pi/2$ radians.
 \item The initial phase $\eta_0$. The prior of $\eta_0$ is set as a uniform distribution between $-\pi$ and $\pi$ radians.
 \item The initial precession angular velocity $\omega_{\rm pre,0}$. As there are at least two precession periods within the $\sim$ 70 seconds duration of the burst, the prior of $\omega_{\rm pre,0}$ is set as a uniform distribution between $0.18$ and $1$ rad/s. Such a range can account for the decrease of $\omega_{\rm pre}$.
 \item The nutation angular velocity $\omega_{\rm nu}$. The prior of $\omega_{\rm nu}$ is set as a uniform distribution between $0.18$ and $2.5$ rad/s.
 \item The precession angular velocity decay index $\xi$. The prior of $\xi$ is set as a uniform distribution between $0$ and $0.5$.
 \item The offset time $t_{0}$ of precession angular velocity decay. The prior of $t_{0}$ is set as a uniform distribution between $-50$ and $10$ seconds.
 \item The characteristic time scale $t_{\rm C}$ of precession angular velocity decay. The prior of $t_{\rm C}$ is set as a uniform distribution between $0$ and $15$ seconds.
 \item The peak time $t_{\rm m}$ of the intrinsic FRED profile. The prior of $t_{\rm m}$ is set as a uniform distribution between $0$ and $10$ seconds.
 \item The rise power-law exponent $r$ of the intrinsic FRED profile. The prior of $r$ is set as a uniform distribution between $0.3$ and $5$.
 \item The decay power-law exponent $d$ of the intrinsic FRED profile. The prior of $d$ is set as a uniform distribution between $0.3$ and $5$.
 \item The peak flux $F_{\rm m}$ of the intrinsic FRED profile. The prior of $F_{\rm m}$ is set as a uniform distribution between $3000$ and $7000$ cts/s.
 \item The time offset $\Delta t$ of the model light curve. The prior of $\Delta t$ is set as a uniform distribution between $-2$ and $2$ seconds.
\end{itemize}

We then performed the fit using a self-developed Bayesian Monte-Carlo fitting package \emph{McEasyFit} \citep{2015ApJ...806...15Z}, which is based on the widely used \textit{Multinest} algorithm \citep{2008MNRAS.384..449F,2009MNRAS.398.1601F}. This package can explore the complete parameter space efficiently to find the reliable best-fit parameters and determine their uncertainties realistically by the converged Markov Chains. The log-likelihood function can be calculated as:

\begin{equation}
 LL =\ln L(M\vert P) =\displaystyle-\frac{1}{2}\sum_{i=1}^n\left[\frac{F(t_{{\rm obs},i}) - M(t_{{\rm obs},i},P)}{\sigma_{i}}\right]^2,
 \end{equation}
where $t_{{\rm obs},i}$ and $F(t_{{\rm obs},i})$ represent the observed time and count rate of the $i^{\rm th}$ data point of the light curve. $F(t_{{\rm obs},i})$ is calculated by summing up the count rate of the GBM detectors n6, n7, and b1 between 10 and 1,000 keV at $t_{{\rm obs},i}$ with a bin size of 0.25 s (blue curve in Figure \ref{fig:precess_model_fit}b), $M(t_{{\rm obs},i},P)$ is the model flux calculated at $t_{{\rm obs},i}$, $\sigma_i$ is the error of $F(t_{{\rm obs},i})$ estimated using the Poisson parameter confidence interval \citep{1986ApJ...303..336G}. Furthermore, the model is required to reproduce two significant peaks in Episode III. Such a condition is guaranteed by forcing $LL$ to be $-\infty$ whenever $M(t_{\rm obs},P) < 3\Bar{\sigma}$ during the time intervals of $61.3 \, {\rm s}\leq t_{\rm obs} \leq 62.3 \, {\rm s}$ and $72.0 \,{\rm s}\leq t_{\rm obs} \leq 72.5 \,{\rm s}$, where $\Bar{\sigma} = 96.8$ cts/s is the variance of the observed background rate.

\subsection{The Result}
\label{subsec:fit_result}

The best-fit model and the corner plot of the posterior probability distributions of the parameters are shown in Figure \ref{fig:precess_model_fit}. Our model successfully fits the observed light curve with PGSTAT/dof = 809.2/626. The best-fit parameters are listed in Table \ref{tab:best_fit}. Our results suggest that the precession-nutation model can well explain the main features of the observed light curve and point to an intrinsic FRED shape emission with $r$ = $1.59_{-0.04}^{+0.14}$, $d$ = $2.22_{-0.09}^{+0.01}$, $t_{\rm m}$ = $5.76_{-0.11}^{+0.13}$ seconds and $F_{\rm m}$ = $4743.38_{-62.15}^{+48.34}$ cts/s produced by a precessing jet with an initial precession period of \textcolor{black}{$18.4\pm0.2$} seconds and a nutation period of \textcolor{black}{$11.1\pm0.2$} seconds. Such an intrinsic shape was 
 modulated to be a periodic-like and missing pattern as observed in GRB 220408B. 

Based on the best-fit parameters in Table \ref{tab:best_fit} and Eq. \ref{equ:openangle}, we can calculate that the observer angle $\theta$ is always smaller than the jet's half-opening angle $\theta_{\rm jet}$, suggesting that the change of $\theta$ does not dominate the change of the laboratory frame observed flux. The lower limit of the jet's half-opening angle can be derived as $\theta_{\rm jet,lolim} = \lambda_{\rm max} + i_{\rm max} = 1.05 \times 10^{-3}$ rad. On the other hand, the precession-nutation effect modulates the shape of the observed light curve mainly through the conversion of photons' observed time between the laboratory frame and the observer frame (i.e., Eq. \ref{equ:time_convert}) rather than the direct influence to the laboratory frame intrinsic light curve (i.e., Eq. \ref{equ:fluxratio}). As a result of the conversion of arrival time between the two frames, the number of arriving photons is redistributed in the observer frame. At certain times, the arrival of photons is more concentrated, which results in the peak structures in the light curve. 

\begin{table}
\setlength\tabcolsep{1pt} 
\centering
\caption{The Best-Fit Parameters of the Light Curve in 10-1,000 keV Range.}
\label{tab:best_fit}
\begin{tabular}{cccc}
\hline
\hline
Parameter&Range&Best-Fit\\
\hline
$\theta_{\rm jet} \,({\rm rad})$&$(0, 0.5]$&$0.33_{-0.22}^{+0.16}$\\
$\lambda_{0} \,({\rm rad})$&$[0, \pi/2]$&$6.45_{-0.06}^{+0.03} \times 10^{-4}$\\
$i \,({\rm rad})$&$[0, \pi/2]$&$3.09_{-0.07}^{+0.02} \times 10^{-4}$\\
$\eta_0 \,({\rm rad})$&$[-\pi, \pi)$&$3.05_{-0.06}^{+0.01}$\\
$\omega_{\rm pre,0} \,({\rm rad/s})$&$[0.18, 1.0]$&$0.34_{-0.01}^{+0.01}$\\ 
$\omega_{\rm nu} \,({\rm rad/s})$&$[0.18, 2.5]$&$0.57_{-0.003}^{+0.0}$\\ 
$\xi \,$&$[0, 0.5]$&$0.32_{-0.01}^{+0.01}$\\
$t_{0} \,\rm(s)$&$[-50, 10]$&$0.28_{-0.88}^{+0.37}$\\
$t_{\rm C} \,\rm(s)$&$(0, 15]$&$11.26_{-0.31}^{+0.94}$\\
$t_{\rm m} \,\rm(s)$&$(0, 10]$&$5.76_{-0.11}^{+0.13}$\\ 
$r$&$[0.3, 5.0]$&$1.59_{-0.04}^{+0.14}$\\ 
$d$&$[0.3, 5.0]$&$2.22_{-0.09}^{+0.01}$\\ 
$F_{\rm m} \,\rm (cts/s)$&$[3000, 7000]$&$4743.38_{-62.15}^{+48.34}$\\ 
$\Delta t \,\rm(s)$&$[-2.0, 2.0]$&$-1.26_{-0.20}^{+0.02}$\\ 
\hline
\end{tabular}
\end{table}

\section{Summary and Discussion}
\label{sec:result_summary}

This {\it Letter} proposes that the observed three-episode feature of GRB 220408B can be explained by a precessing jet. Based on the similarities between the three episodes in light curve profile, spectral evolution, and spectral lags, we concluded that they may have the same origin and may be the result of jet precession. A jet-precession model can be successfully used to fit the light curve of GRB 220408B, which assumes a FRED shape light curve that precesses and nutates with slowing precession angular velocity. Our fit suggests that the photon arrival time change in different frames resulting from the precession jet plays a prominent role in shaping the observed light curve when a GRB is observed off-axis.

In view of computation costs, our model, which already has 14 free parameters, does not incorporate the reproduction of the observed spectral evolution. Although, in principle, such evolution can be attributed to the Doppler factor change predicted by our model, a realistic model should also take into account the intrinsic central engine behaviors that result in the observed spectral evolution, which adds even more complexity to the model but will be left for future study. 

\section*{acknowledgements}

We acknowledge support by the National Key Research and Development Programs of China (2018YFA0404204,2022YFF0711404,2022SKA0130100, 2022SKA0130102, 2018YFA0404502), the National Natural Science Foundation of China (Grant Nos. 11833003, U2038105, 12121003, 12025301 \& 11821303), the science research grants from the China Manned Space Project with NO.CMS-CSST-2021-B11, the Program for Innovative Talents, Entrepreneur in Jiangsu. Y.-Z.M. is supported by the National Postdoctoral Program for Innovative Talents (grant no. BX20200164). We acknowledge the use of public data from the Fermi Science Support Center (FSSC).

%\bibliography{ms.bib}
% \bibliography{ms}{}
% \bibliographystyle{aasjournal}

\end{document}